\documentclass[aps, prl, superscriptaddress, reprint,nobibnotes]{revtex4-2}
\usepackage{amsmath}
\usepackage{color}
\usepackage{comment}
\usepackage{bm,graphicx,hyperref}
\hypersetup{%
  breaklinks = {true},
  citecolor = {blue},
  colorlinks = {true},
  linkcolor = {red},}

\usepackage{lineno}
\usepackage[caption=false]{subfig}
\usepackage[bottom]{footmisc}
\usepackage{color,soul}
\usepackage{newfloat}

\DeclareFloatingEnvironment[
    fileext=los,
    listname={List of Schemes},
    name=SCHEME,
    placement=tbhp,
    within=section,
]{scheme}

\begin{document}
\title{Gate-tunable artificial nucleus in graphene}

\author{Mykola Telychko}
\thanks{These authors contributed equally to this work.}
\affiliation{Department of Chemistry, National University of Singapore,  117543, Singapore}

\author{Keian Noori}
\thanks{These authors contributed equally to this work.}
\affiliation{Centre for Advanced 2D Materials (CA2DM), National University of Singapore, 117543, Singapore}

\author{Hillol Biswas}
\affiliation{Department of Physics, National University of Singapore, 2 Science Drive 3, 117542, Singapore}
\affiliation{Centre for Advanced 2D Materials (CA2DM), National University of Singapore, 117543, Singapore}

\author{Dikshant Dulal}
\affiliation{Yale-NUS College, 16 College Avenue West,  138527, Singapore}

\author{Pin Lyu}
\affiliation{Department of Chemistry, National University of Singapore, 117543, Singapore}

\author{Jing Li}
\affiliation{Centre for Advanced 2D Materials (CA2DM), National University of Singapore, 117543, Singapore}

\author{Hsin-Zon Tsai}
\affiliation{Department of Physics, University of California, Berkeley, California 94720, USA.}

\author{Hanyan Fang}
\affiliation{Department of Chemistry, National University of Singapore,  117543, Singapore}

\author{Zhizhan Qiu}
\affiliation{Department of Chemistry, National University of Singapore,  117543, Singapore}

\author{Zhun Wai Yap}
\affiliation{Yale-NUS College, 16 College Avenue West,  138527, Singapore}

\author{Kenji Watanabe} 
\affiliation{Research Center for Functional Materials,
National Institute for Materials Science, 1-1 Namiki, Tsukuba 305-0044, Japan}

\author{Takashi Taniguchi}
\affiliation{International Center for Materials Nanoarchitectonics,
National Institute for Materials Science,  1-1 Namiki, Tsukuba 305-0044, Japan}

\author{Michael F. Crommie}
\affiliation{Department of Physics, University of California, Berkeley, California 94720, USA.}

\author{Aleksandr Rodin}
\thanks{Corresponding author}
\affiliation{Yale-NUS College, 16 College Avenue West,  138527, Singapore}
\affiliation{Centre for Advanced 2D Materials (CA2DM), National University of Singapore, 117543, Singapore}

\author{Jiong Lu}
\thanks{Corresponding author}
\affiliation{Department of Chemistry, National University of Singapore, 117543, Singapore}
\affiliation{Centre for Advanced 2D Materials (CA2DM), National University of Singapore, 117543, Singapore}
\affiliation{Institute for Functional Intelligent Materials, National University of Singapore, 117544, Singapore}
\date{\today}

\begin{abstract}

We report an atomically-precise integration of individual nitrogen (N) dopant as an in-plane artificial nucleus in a graphene device by atomic implantation to probe its gate-tunable quantum states and correlation effects. The N dopant creates the characteristic resonance state in the conduction band, revealing a giant carrier-dependent energetic renormalization up to 350 meV with respect to the Dirac point, accompanied by the observation of long-range screening effects. Joint density functional theory and tight-binding calculations with modified perturbation potential corroborate experimental findings and highlight the short-range character of N-induced perturbation.

\end{abstract}	
\maketitle

The relativistic-like electronic dispersion in graphene offers a rich platform for studying exotic quantum electrodynamics phenomena in a condensed-matter setup, as well as for developing novel electronics and quantum optics ~\citep{Ni2018, Fei2012}. This photon-like band structure also makes graphene electrons much less amenable to electrostatic confinement compared to conventional materials. A peculiar type of electron confinement in graphene arises from the presence of strong Coulomb charges in the so-called supercritical regime, analogous to the supercritical collapse of super-heavy nuclei in quantum electrodynamics ~\citep{Wang734, Pomeranchuk1945}. Such a degree of local density enhancement has been achieved by the assembly of highly charged transition metal adatoms~\citep{Wong2017, Wang2012, Wang2012a, Brar2011}, one-dimensional molecular arrays on graphene~\citep{Tsai2020, Lu2019}, and creating a charged vacancy in graphene~\citep{Mao2016}. The extent of the potential produced by ``out-of-plane" charge centers above graphene is expected to be very different from the ``in-plane" charges hosted in graphene.~\citep{Pereira2007, Shytov2007, Shytov2007a}. This distinction hinges on a question: Is it possible to fabricate atomic charges in a graphene lattice that can behave as a robust artificial nucleus in a \textit{true} 2D limit?

The ideal scenario of an ``in-plane" Coulomb charge center assumes a point charge residing in the graphene plane without imposing any lattice deformations. The most viable way to create such a scenario is to utilize nitrogen (N) substitution into graphene. The similarity of atomic radii of N and C, accompanied by the ability of N to adopt $sp_2$-hybridization, allows for its facile incorporation into the graphene lattice. In addition, due to its proximity to carbon in the periodic table, $sp_2$-like N substitution mimics the inclusion of an extra proton into the graphene lattice. Both atom-resolved imaging ~\cite{Telychko2014, Telychko2015,Lagoute2015, GonzalezHerrero2016asc, Mallada2020, Zhao2011, Wang2012b, Joucken2019, VanDerHeijden2016,Velasco2021} and sample-averaged spectroscopic techniques ~\cite{Sforzini2016, Tripathi2018ebm,  DelaTorre2018, Lin2019, Bouatou2020} have been deployed to probe structural and electronic properties of N impurities in graphene. Nevertheless, all these studies utilized heavily-doped graphene layers with altered intrinsic electronic structure due to the presence of the underlying substrates, including silicon-carbide~\cite{Telychko2014, Joucken2019}, Cu(111)~\cite{Zhao2011, VanDerHeijden2016}, or another graphene layer~\cite{Velasco2021}, that hamper the deciphering of both short- and long-range electronic effects associated with N atoms in pristine single-layer graphene at varied charge carrier densities. Therefore, back-gated graphene-boron nitride (G/BN) devices harboring atomically smooth and charge-homogeneous interfaces~\cite{Dean2010} may provide an ideal platform to probe unique electronic behavior around N impurities in a tunable and poorly-screened environment.

\begin{figure}
    \centering
    \includegraphics[width=\columnwidth]{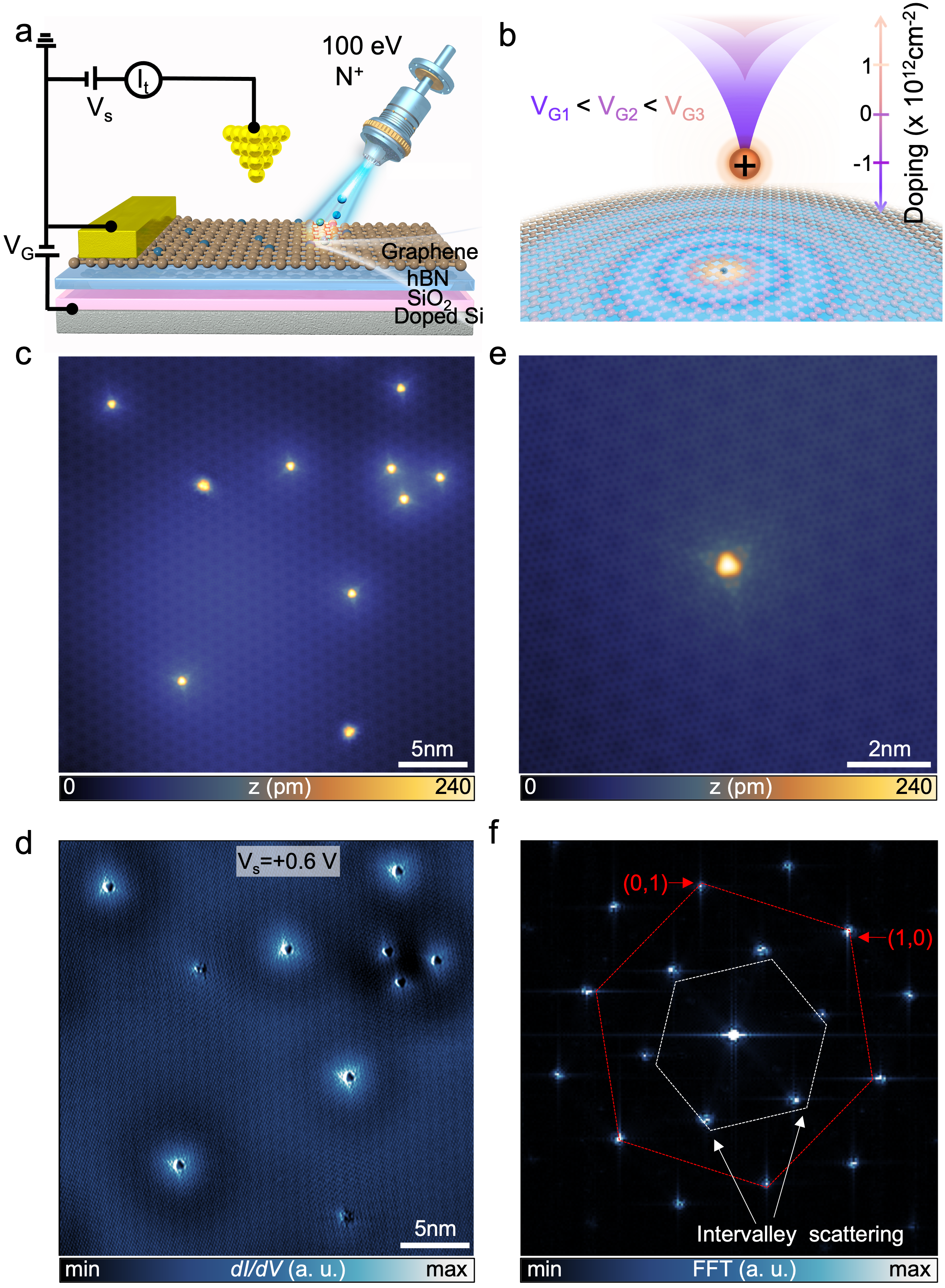}
    \caption{\textbf{Creating substitutional nitrogen dopants in back-gated G/BN}. (a) Schematics showing the route towards incorporation of N dopant into back-gated G/BN using low-energy (+100 eV) N$^+$ ion implantation. (b) Schematic illustration of a tunable proton-induced potential \textit{via} gating. (c) Large-scale STM image reveals the presence of individual substitutional N atoms. (d) \textit{dI/dV} map acquired at $V_s=+0.6$ V over the same surface area as STM image in panel (c). (e) Zoom-in image of individual N dopant. (f) Fast Fourier transformation of the STM image of individual N dopant.}
    \label{fig:Device}
\end{figure}

Here, we have devised a well-controlled N$^+$ ion implantation routine allowing for a high-yield $sp_2$-like incorporation of individual N atoms into the graphene lattice (Fig.~\ref{fig:Device}a,b) in a back-gated G/BN device. Individual N dopants behave as in-plane artificial protons, manifested by the emergence of a broad resonance state ($E_N$) above the Dirac point. Gate-dependent scanning tunneling spectroscopy (STS) measurements reveal a giant energetic shift of the $E_N$ with respect to the Dirac point together with long-range Friedel oscillations in the vicinity of N dopants. Density functional theory (DFT) and effective tight-binding (TB) theory calculations with modified on-site perturbation potential and hopping constant corroborate the emergence of the new resonance state and its associated gate-tunable correlation effects.

We first fabricated the back-gated device consisting of CVD-grown monolayer graphene on top of a hexagonal boron-nitride flake (G/BN) placed onto a SiO$_2$ substrate, as shown in Fig.~\ref{fig:Device}a (also Fig.~\ref{back_device}). Gate voltage ($V_G$), applied between the electrostatically-grounded graphene sheet and the doped Si wafer, enables the tunability of the carrier density in graphene (Fig.~\ref{fig:Device}b). 

Our route towards the fabrication of N-based artificial nuclei in back-gated G/BN (Fig.~\ref{fig:Device}a) relies on its posterior bombardment using low-energy (+100 eV) N$^+$ ions in ultra-high vacuum conditions, as illustrated in Fig.~\ref{fig:Device}a (see Methods for further details). This process results in a robust $sp_2$-like incorporation of individual N atoms into graphene. The representative STM images reveal that the surface of the G/BN after N$^+$ implantation is decorated with abundant triangular-shaped features attributed to the individual N dopants (Fig.~\ref{fig:Device}c).

High-resolution STM imaging (Fig.~\ref{fig:Device}e) and $dI/dV$ mapping (Fig.~\ref{fig:Device}d) of the single N dopant in graphene demonstrate the characteristic $\sqrt{3}\times\sqrt{3}$ pattern associated with the intervalley electron scattering processes, which has also been revealed in the vicinity of atomic-scale defects, including hydrogen adatoms~\cite{Dutreix2019} and carbon vacancies~\cite{Ugeda2010, Mao2016}. Characteristic spots attributed to electron back-scattering processes between K and K$^\prime$ valleys are also evident in the Fourier-transformed STM image (Fig.~\ref{fig:Device}f). 
\begin{figure}
    \centering
    \includegraphics[width = \columnwidth]{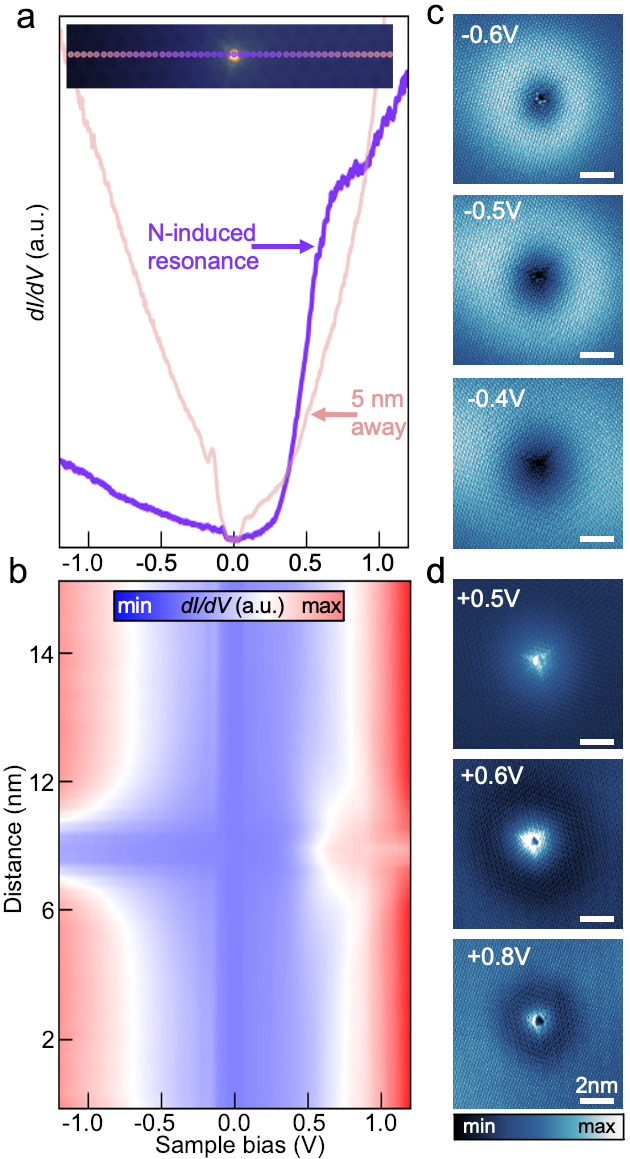}
    \caption{\textbf{Characterization of individual N dopants}. (a) \textit{dI/dV} spectra acquired over N-site and pristine graphene. (b) 2D color-coded contour plot of the \textit{dI/dV} spectra acquired along the line traversing single N dopant indicated in the inset STM image in panel (a). (c,d)  \textit{dI/dV} maps acquired at negative (c) and positive (d) bias voltages indicated in top-left corners. Lengths of all scale bars is 2 nm.} 
    \label{fig:Electronic_Structure_N}
\end{figure}
\begin{figure}
    \centering
    \includegraphics[width = \columnwidth]{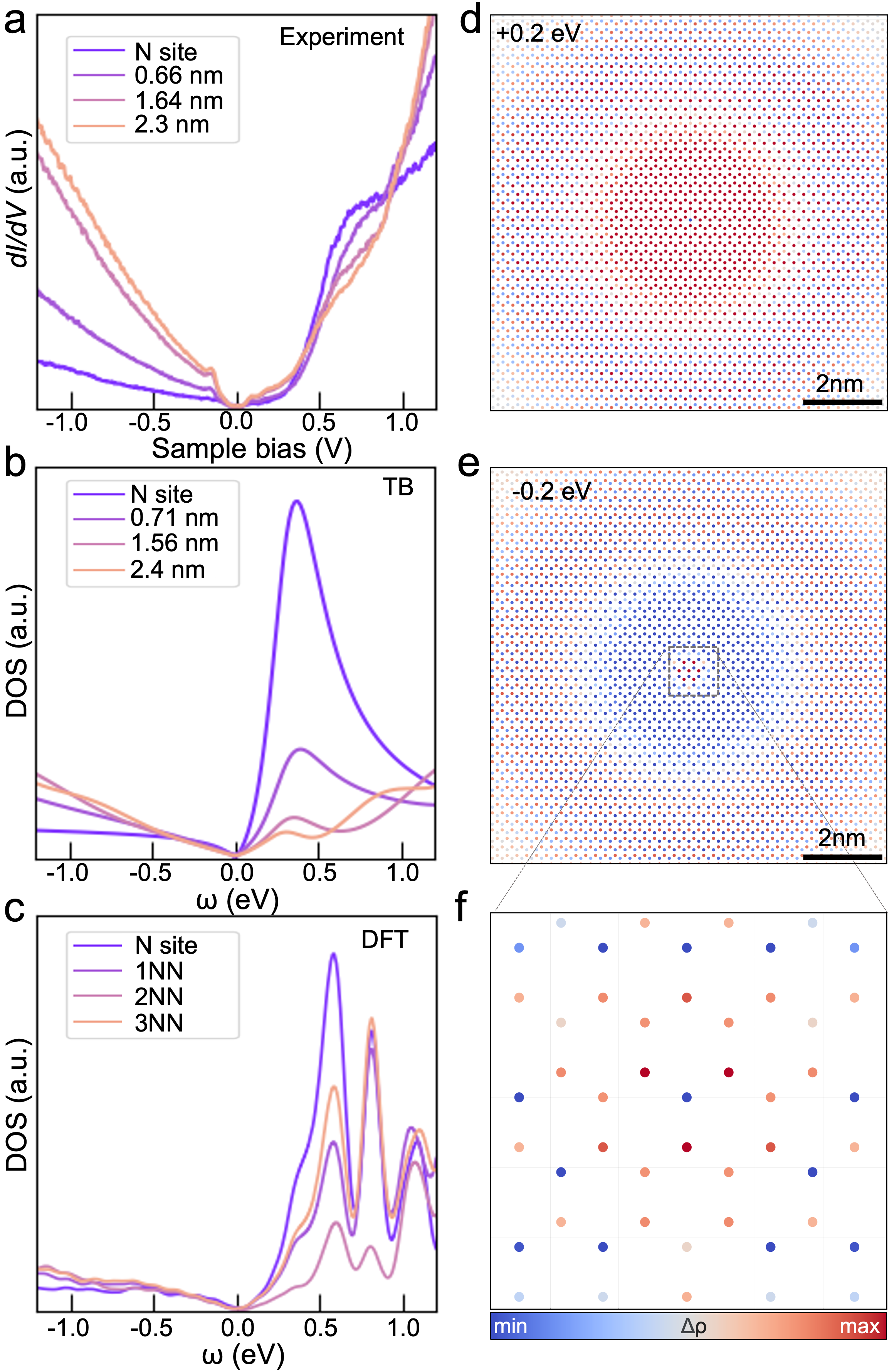}
    \caption{\textbf{Electronic structures of individual N dopants}. (a) \textit{dI/dV} spectra acquired over an N site and its vicinity. (b) Site-specific density of electronic states (DOS) spectra acquired over an N site and its vicinity using TB model for $\Delta_0 = -7$ eV, $\Delta_1 = -0.7$ eV, and $\delta t = 0.5$ eV. (c) DFT-calculated DOS of the N site and its first (1NN), second (2NN) and third (3NN) nearest C neighbours in a $10\times10$ supercell. (d,e) Real space spectral maps calculated at energies of +0.2 eV (d) and -0.2 eV (e) using TB formalism. (f) Zoom-in image of area marked by square in panel (e).}
    \label{TB_DFT}
\end{figure}

\begin{figure*}
    \centering
    \includegraphics[width=15cm]{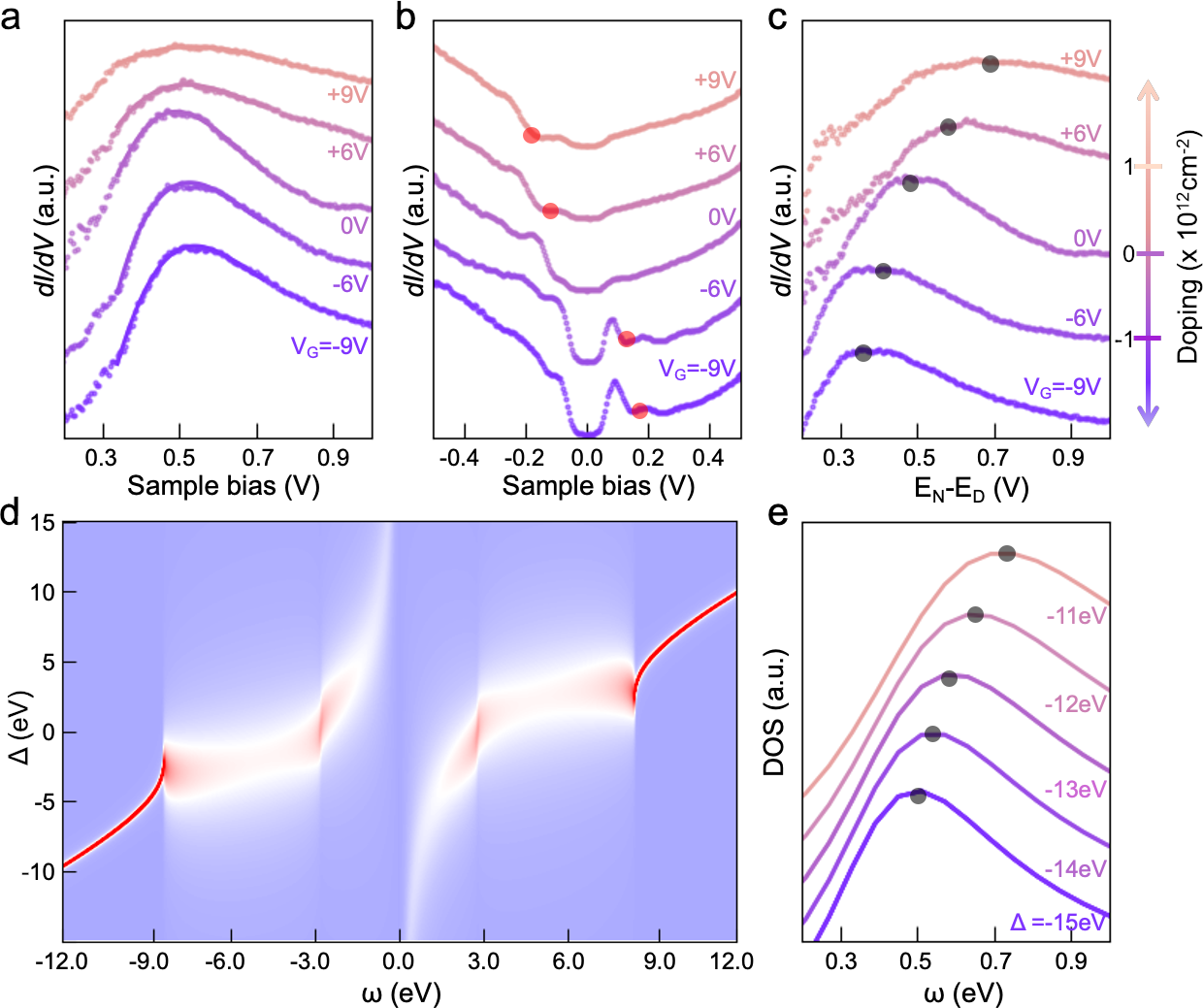}
    \caption{\textbf{Gate-tunable electronic properties of N dopant}. (a) \textit{dI/dV} spectra acquired over N site at various $V_G$. (b)  \textit{dI/dV} spectra acquired over pristine graphene area at various $V_G$. Red dots mark positions of Dirac point ($E_D$). (c) Energetic positions of N-induced resonance state ($E_N$) with respect to $E_D$ extracted from site-specific \textit{dI/dV} spectra acquired at various $V_G$.(d) Color-coded TB spectral plot. (e) TB spectra extracted for monotonously varied values of the perturbation potential ($\Delta$).}
    \label{Gate-tunable}
\end{figure*}

Next, we probed the electronic properties of the N dopants \textit{via} the acquisition of the site-specific $dI/dV$ point spectra in the nearly-neutral doping regime (i.e., $V_G=0$ V), which reveal several prominent features (Fig.~\ref{fig:Electronic_Structure_N}a,b). These include (i) a gap-like feature ($\sim 130$ meV) centered at the Fermi level ($E_F$) attributed to the phonon-assisted inelastic tunneling~\cite{Zhang2008}; (ii) an electron-hole asymmetry with reduced (increased) spectral weight of the hole (electron) $dI/dV$ branch, decaying with distance from the N site (Fig.~\ref{TB_DFT}a). This indicates a net positive charge of the N dopant ~\cite{Mallada2020, Wong2015a} as expected from a partially screened proton in the graphene lattice; (iii) the $dI/dV$ spectra acquired over the N site (Fig.~\ref{fig:Electronic_Structure_N}a) and in its proximity reveal a broad resonance state (denoted as $E_N$) centered at $\sim 590$ meV above $E_F$, as further illustrated by the normalized $dI/dV$ spectra (Fig.~\ref{normalised}). The 2D contour map of the $dI/dV$ spectra (Fig.~\ref{fig:Electronic_Structure_N}b) reveals the spatial extent of the N-induced resonance state gradually decaying with distance from the N site and fading beyond $\sim 1.7$ nm. 

In addition, bias-dependent $dI/dV$ maps reveal characteristic concentric rings in the vicinity of the N dopant that are attributed to the Friedel oscillations (FO's). The $dI/dV$ maps acquired at negative sample biases (Fig. ~\ref{fig:Electronic_Structure_N}c) display an attenuated $dI/dV$ intensity in direct proximity to the N site surrounded by brighter FO rings. In contrast, $dI/dV$ maps acquired at positive sample biases (Fig. ~\ref{fig:Electronic_Structure_N}d) feature a reversed contrast with an enhanced $dI/dV$ signal close to the N site, surrounded by FO rings with a lower $dI/dV$ intensity. It is noteworthy that the diameter of the FO rings scales inversely with the bias voltage as revealed also by normalized $dI/dV$ linescans acquired at different $V_s$ shown in ~Fig. ~\ref{didv_profile}b.  For instance, the $dI/dV$ linescans acquired at $V_s$= $-0.6$ V, $-0.5$ V and $-0.4$ V feature FO rings with a diameter of $4.16$ nm, $5.76$ nm, $7.30$ nm, respectively.  The FO's diameter is also found to be tunable by charge-carrier density as evidenced by $dI/dV$ linescans acquired at varied $V_G$ Fig.~\ref{didv_profile}a.

We then explored carrier-dependent electronic properties of the N dopant by acquiring the $dI/dV$ spectra at various $V_G$ ranging from $V_G=-9$ V (p-doped regime) to $V_G=+9$ V (n-doped regime). Prior to this, gate-dependent $dI/dV$ spectra acquired over nearby graphene ($> 10$ nm away) exhibit a characteristic energy shift of Dirac point (``dip-feature" marked by red dots) in Fig.~\ref{Gate-tunable}b, whereby the carrier concentration at $V_G=+9$ V is estimated to be $n_e\approx 2.1 \times 10^{12} $ cm$^{-2}$ using a standard capacitor model ~\citep{Qiu2019,Decker2011} (see Methods). Note that the charge neutrality point of graphene is close to $V_G$ = 0 (Fig.~\ref{si_gate}). In contrast, the normalised point $dI/dV$ spectra (Fig.~\ref{normalised}) acquired over the N site reveal a slightly $V_G$-dependent energetic shift of $E_N$, from $497$ mV (at $V_G=-9$ V) to $464$ mV (at $V_G=0$ V) (Fig.~\ref{Gate-tunable}a). Further, gating graphene into the n-doped regime ($V_G=+9$ V) leads to a notable broadening of $E_N$ and an attenuation of its spectral weight.

The site-specific $dI/dV$ measurements discussed above enable the determination of the $V_G$-dependent energetic position of the $E_N$ with respect to the Dirac point  (Fig.~\ref{Gate-tunable}c). It is evident that a gradual variation of the back-gate voltage from $V_G=-9$ V to $V_G=+9$ V leads to a monotonic shift of $E_N$ from $\sim 350$ mV to ~$\sim 700$ mV relative to the Dirac point (black dots in Fig.~\ref{Gate-tunable}c). Such a giant carrier density-dependent energetic renormalization of the N-induced resonance state with respect to the Dirac point sets N dopants apart from other previously studied impurities residing in back-gated graphene. Specifically, it has been shown that the electronic resonance states associated with hydrogen adatoms,~\cite{GonzalezHerrero2016asc, Noori2020}, single carbon vacancies, and ~\cite{Mao2016} cobalt clusters~\cite{Wang2012} are energetically pinned to the Dirac point and thus exhibit a rigid shift with the Dirac point as a function of $V_G$. The non-rigid shift of molecular frontier orbitals with respect to $V_G$ has been recently attributed to many-electron interactions and gate-tunable graphene's polarization effects ~\cite{Wickenburg2016}.

We first carried out DFT calculations to understand both local structural and electronic properties of a single N dopant in graphene. Specifically, DFT-calculated projected density of states (PDOS) for the $p_z$ orbitals of the N dopant, along with its first, second, and third nearest C neighbors, reveals a peak centered at $\sim 500$ meV above $E_F$ (Fig.~\ref{TB_DFT}c), resembling the features observed experimentally. In addition, our DFT calculations also indicate that the presence of N in graphene renders a rather negligible lattice distortion, with almost identical lengths of C-N (1.41~\AA) and C-C (1.42~\AA) bonds. Moreover, graphene's lattice remains flat, retaining its $sp_2$ hybridization, preventing any spin-flip processes, and preserving the separation of graphene $\pi$ and $\sigma$ bands. Consequently, N substitution allows for the formation of the desired in-plane Coulomb center while maintaining the $sp_2$ integrity of graphene. 

To describe both near- and far-field behavior around the N dopant, we employ tight-binding (TB) calculations with the formalism described in Ref.~\citep{Noori2020a} using a Julia language~\cite{Bezanson2017} package~\cite{GrapheneQFT}. Since the TB approach with nearest-neighbor hopping works well for graphene, and since the $p_z$ orbitals of nitrogen are smaller than those of carbon (Fig.~\ref{FigS8}), it should be possible to only change the on-site energy of the dopant and its neighbors (by adding energies $\Delta_0$ and $\Delta_1$, respectively) and adjust the hopping parameters ($\delta t$) to describe the N-induced broad resonance state. Indeed, we find that such a perturbation, with $\Delta_0 = -7$ eV, $\Delta_1 = -0.7$ eV, and $\delta t = 0.5$ eV, shown in Fig.~\ref{TB_DFT}b, gives a good agreement with the \emph{ab initio} results (Fig.~\ref{TB_DFT}c), reinforcing the short-range nature of the N-induced perturbation. In addition, real-space spectral function maps calculated using the TB formalism reproduce experimental $dI/dV$ maps by revealing both the characteristic triangular shape near to N and the surrounding FO's concentric rings (Fig. \ref{TB_DFT}d,e).

As discussed above, substituting a C with an N can be viewed as injecting a proton into the system. Consequently, one might expect the emergence of long-range effects arising from the Coulomb potential produced by this proton instead of the short-range perturbation suggested by the DFT and TB calculations. To reconcile these two pictures, one can start with the proton-generated $1 / r$ attractive potential, which gives rise to bound states with energies below the lowest kinetic energy that particles can have. In the case of graphene, the relevant system of particles corresponds to the $\pi$-band electrons. DFT calculations confirm the existence of a bound state below the $\pi$ band (Fig.~\ref{FigS10}). Naturally, this state will become occupied, and the electron will screen the proton, sharply diminishing the extent of the Coulomb potential. Additionally, the surrounding atoms also screen the positive charge, further reducing the potential's range.

One last question that remains is the gate-dependent renormalization of the N-induced resonance state ($E_N$) with respect to the Dirac point. A variation of carrier density in graphene is expected to further modulate the screened potential of the added proton, equivalent to tuning the on-site energy of the dopant and its neighbors, and the hopping parameter between them, in the TB formalism.

More specifically, sweeping the back-gate voltage from $V_G=-9$ V to $V_G=+9$ V corresponding to an electron density of $n_e\approx 2.1 \times 10^{12} $ cm$^{-2}$ in the system. As a result, this provides additional screening to the already-screened proton, effectively weakening its ability to lower the potential energy of other electrons. In our toy model, this is equivalent to making on-site energies less negative and pushing the resonance state to higher energies (Fig.~\ref{Gate-tunable}d). While the entire band structure shifts to lower energy due to gating, the net effect is manifested as the $V_G$-dependent $E_N$ resonance state lagging behind the Dirac point (Fig.~\ref{Gate-tunable}c). As the simplest illustration, we only include the on-site energy modification $\Delta$ for the nitrogen dopant and plot the dopant's spectral function for a range of $\Delta$'s in Fig.~\ref{Gate-tunable}d. The respective plots for the (next-)nearest carbon neighbor are shown in Fig.~\ref{FigS11}. We observe that as the on-site potential is varied, the energy difference between the resonance in the conduction band and the Dirac point changes (Fig.~\ref{Gate-tunable}e), as seen in the experiment. In addition, one can expect the formation of a bound state below the band minimum for sufficiently large $\Delta$'s, as seen in Fig.~\ref{Gate-tunable}d.

In summary, we have demonstrated a robust technique towards a controllable  $sp_2$-like incorporation of N atoms into back-gated Gr/BN. The N dopant, seamlessly embedded into the graphene lattice, behaves as an artificial proton with dominant short-ranged potential perturbation, leading to the formation of broad resonance state above the Dirac point. The strength of the proton's potential can be effectively modified  \textit{via} tuning charge carrier density in graphene, leading to a giant gate-tunable energetic renormalization of the resonance state up to 350 meV with respect to the Dirac point. It can be envisaged that such an atomically-precise N$^+$ implantation technique, combined with a judiciously-designed device patterning strategy, may enable facile fabrication of nanoscale linear ~\citep{Wang2018_pn, Lin2015} or circular ~\citep{Lee2016,ge2020,Ge2021} p-n junctions in graphene for developing novel electronics and quantum optics.

\section{Acknowledgements}
J. Lu acknowledges the support from MOE (Singapore) through the Research Centre of Excellence program (grant EDUN C-33-18-279-V12,I-FIM) and MOE Tier 2 (MOE2019-T2-2-044). M. Telychko acknowledges the support from A*STAR AME YIRG grant (Project No. A20E6c0098, R-143-000-B71-305). A. Rodin acknowledges the National Research Foundation, Prime Minister Office, Singapore, under its Medium Sized Centre Programme and the support by Yale-NUS College (through Grant No. R-607-265-380-121). K. Watanabe and T. Taniguchi acknowledge support from the Elemental Strategy Initiative conducted by the MEXT, Japan (Grant Number JPMXP0112101001) and  JSPS KAKENHI (Grant Numbers 19H05790, 20H00354 and 21H05233).

\section{Author Contributions}
 M.T. A.R., and J.Lu. conceived the project. M.T. performed experiments related ion implantation of Gr/BN, STM/STS measurements and data analysis. K.N, H.B., D. D., Z.W.Y., and A.R. performed DFT and TB calculations. P.L., J.Li. fabricated back-gated Gr/BN device under guidance of H.-Z.T. and M.F.C. K.W. and T.T grew hBN crystals for the device. H.F and Z.Q. assisted in STM experimental procedures. The manuscript was written by M.T., A.R. and J.Lu. with contributions from all co-authors.

%

\newpage
\onecolumngrid
\clearpage
\section{Supplementary Information}
\appendix
\renewcommand{\thefigure}{S\arabic{figure}}
\setcounter{figure}{0}

\section{Experimental methods}

\textbf{Gr/BN device fabrication}. A back-gated graphene/hBN/SiO$_2$ device was prepared by overlaying CVD-grown graphene onto hexagonal boron nitride (hBN) flakes exfoliated onto a SiO$_2$/Si substrate. The thickness of SiO$_2$ layer was 90 nm. 

A value of charge-carrier density was determined using relation $n=\alpha V_G$. Here, $\alpha$ was estimated to be $\sim23.3 \times 10^{10}$ cm$^{-2}$V$^{-1}$ using capacitor model consisting of SiO$_2$ layer with thickness of 90 nm \cite{Novoselov2005, Zhang2005}.  

\textbf{N implantation procedure}. The Gr/BN device was subsequently exposed to a low energy (+100 eV)  N$^+$ ions flux at base pressure 1x10$^{-7}$ mbar for 2 minutes. For the N$_2$ gas ionization we used the conventional filament-based ionization source IQE 11/35 manufactured by Specs.

\textbf{The STM and STS measurements}. The STM experiments were performed in UHV conditions at temperature 4.4 K using a commercial Specs LT STM machine. The STM tips were calibrated on the Au(111) surface by measuring Shockley surface state prior to all STS measurements. The \textit{dI/dV} curves were acquired under open feedback conditions by lock-in detection of an alternating tunnel current with a modulation voltage of 6–16 mV (r.m.s.) at 780 Hz. 

\textbf{DFT calculations}.
DFT calculations were performed using the Quantum \textsc{Espresso} package~\citep{Giannozzi2009,Giannozzi2017} using a PAW basis~\citep{Blochl1994,DalCorso2014} with Perdew-Burke-Ernzerhof (PBE) exchange-correlation functional~\citep{Perdew1996}. The kinetic energy cutoffs for wavefunctions and charge densities were set to at least the minimum recommended values specified in the basis~\citep{DalCorso2014}. The reciprocal Brillouin zone (BZ) was sampled \textit{via} a uniform, unit cell-equivalent grid of $60 \times 60 \times 1$ k-points for charge density calculations, and $120 \times 120 \times 1$ k-points for eigenvalue (DOS) calculations. 

\textbf{TB calculations}. The calculations were performed using GrapheneQFT package~\citep{GrapheneQFT} written in JULIA programming language~\cite{Bezanson2017}.

\clearpage

\begin{figure}
    \centering
    \includegraphics[width = 15cm]{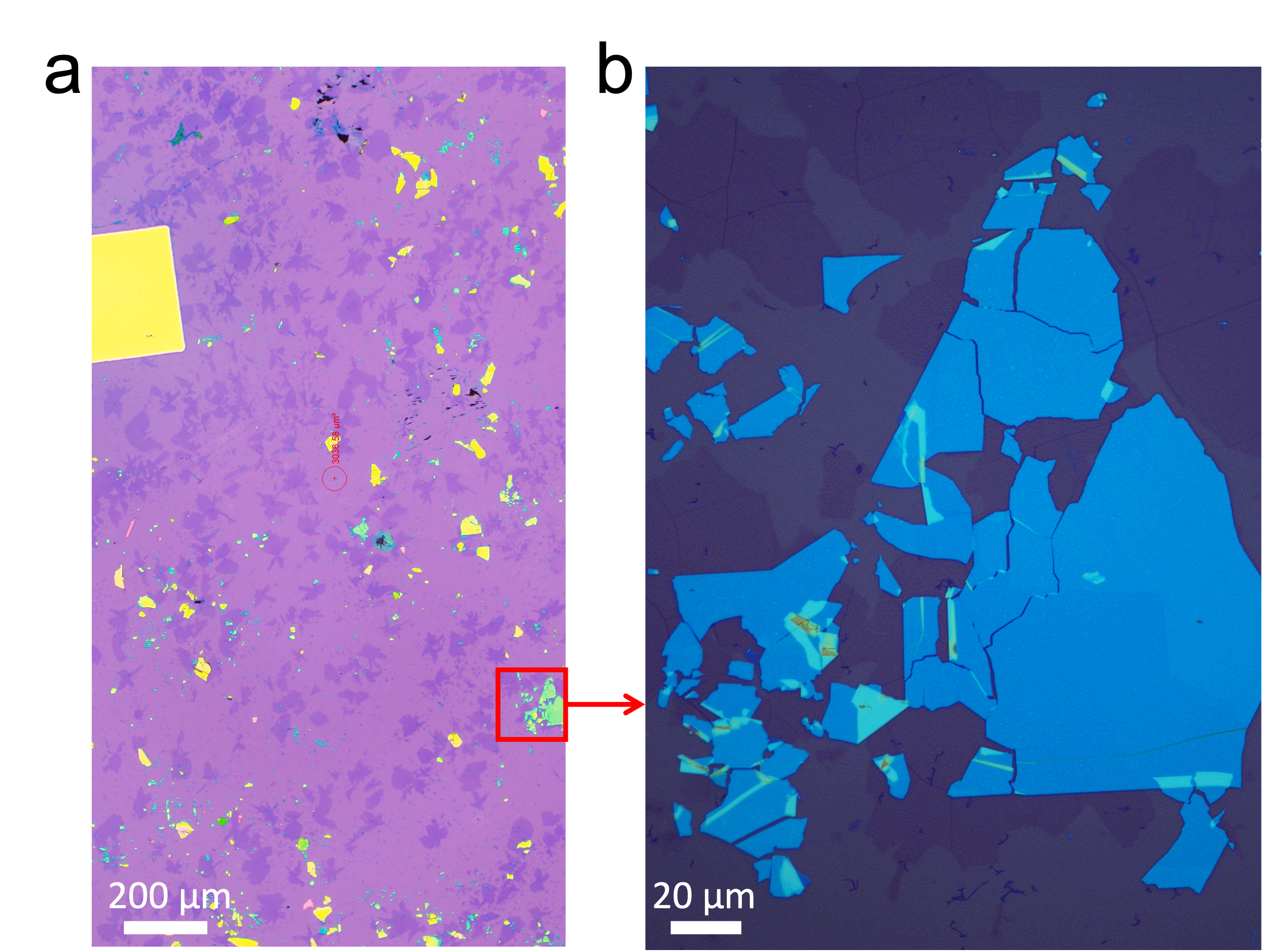}
    \caption{(a,b) The optical images of  back-gated graphene/hBN/SiO$_2$/Si device.}
    \label{back_device}
\end{figure}

\begin{figure}
    \centering
    \includegraphics[width =12cm]{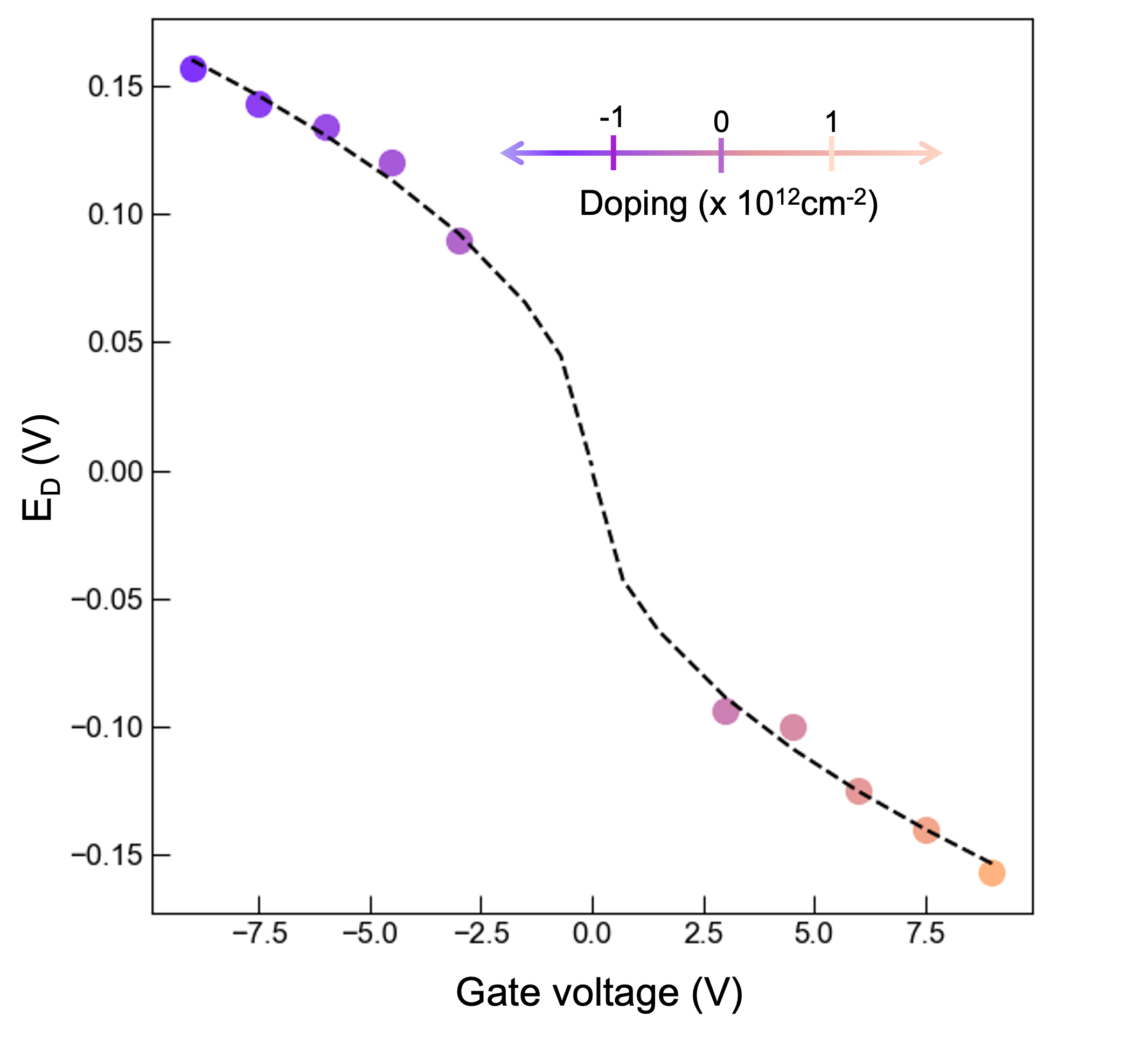}
   \caption{Energetic position of Dirac point ($E_D$) as a function of gate voltage ($E_G$) was extracted from characteristic conductance minima in $dI/dV$ curves acquired at various $V_G$ (see Fig.~\ref{Gate-tunable}b). Dashed line represent result of the $E_D \sim \sqrt{V_G}$ fitting. }
    \label{si_gate}
\end{figure}

\begin{figure}
    \centering
    \includegraphics[width = 20cm]{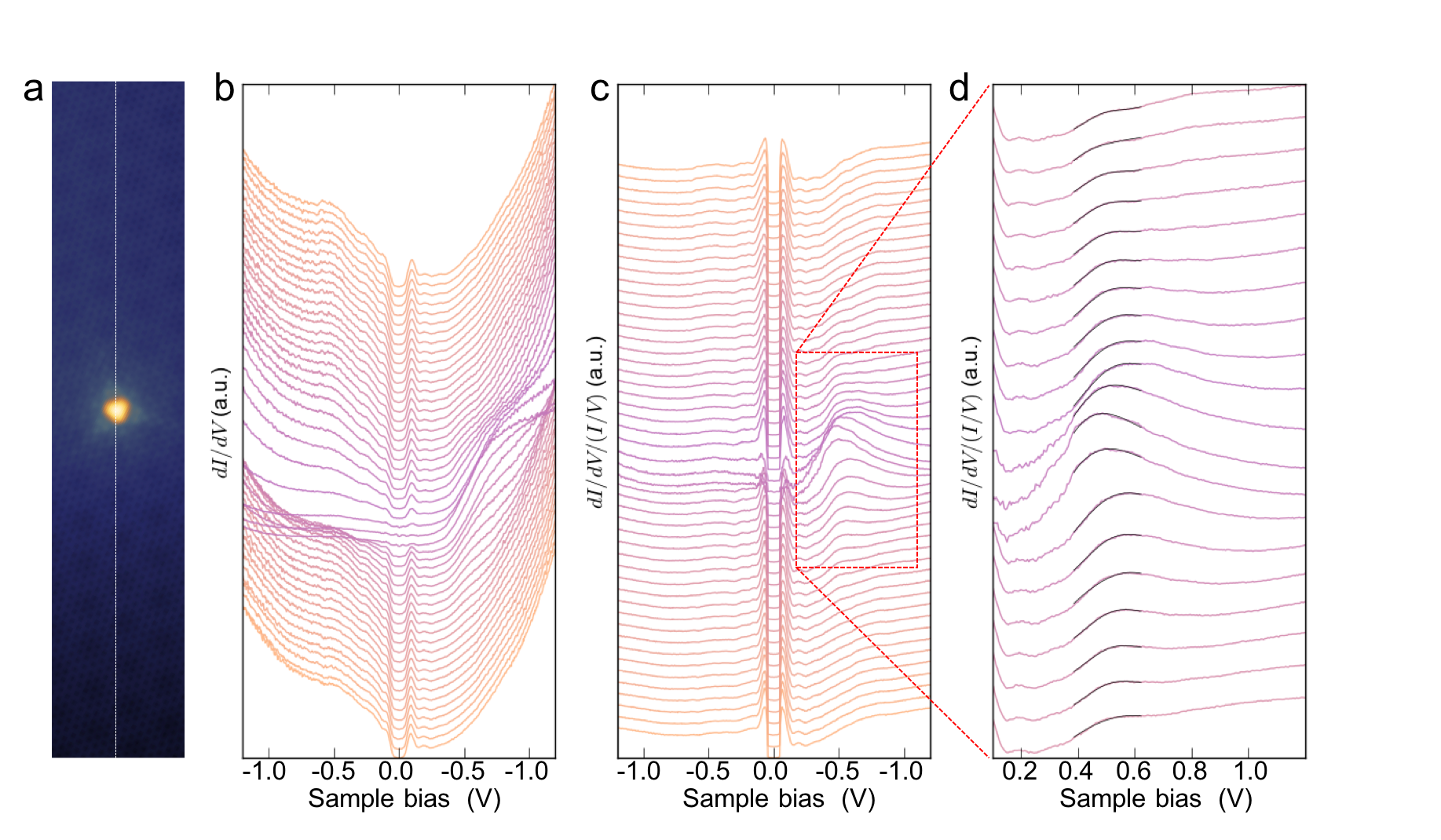}
    \caption{\textbf{Normalisation of the $dI/dV$ spectra}. (a) STM image of individual N dopant. (b) Sequence of point \textit{dI/dV}spectra acquired across N dopant, along white dashed white line shown in panel (a). (c) Normalized $dI/dV$ spectra. (d) Representative fitting result of the N-induced resonance state using Lorentzian function denoted by black curves.}
    \label{normalised}
\end{figure}

\begin{figure}
    \centering
    \includegraphics[width = 15cm]{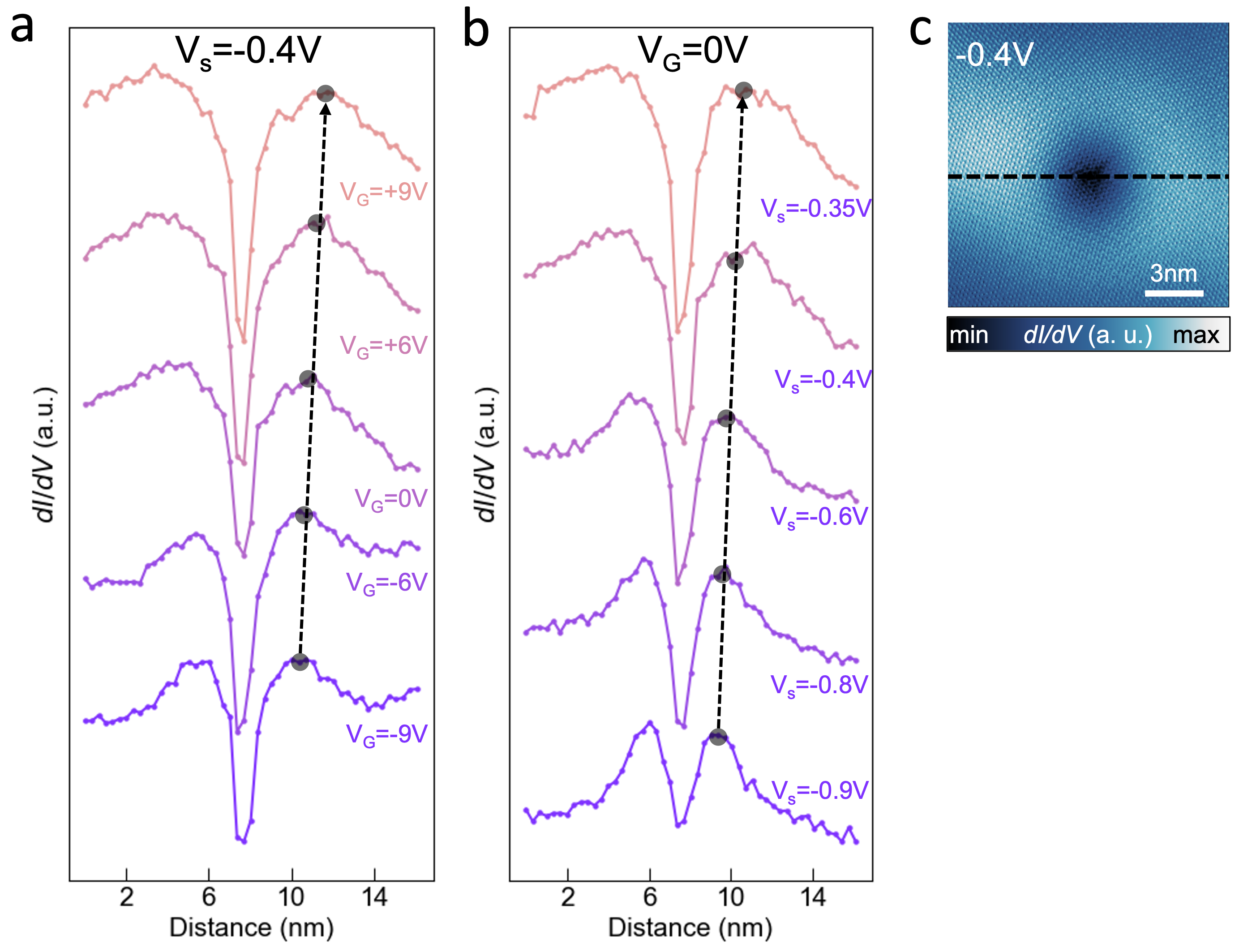}
    \caption{\textbf{$dI/dV$ versus distance linescans traversing N dopant}. (a) Normalized $dI/dV$ linescans acquired at $V_s$=$-0.4$ V at varied $V_G$. (b) Normalized $dI/dV$ linescans acquired at varied $V_s$ and $V_G$=0 V. (c) $dI/dV$ map acquired at $V_s$=$-0.4$. Black dashed line indicates direction of $dI/dV$ profile in panels (a) and (b).}
    \label{didv_profile}
\end{figure}

\begin{figure}
    \centering
    \includegraphics[width =18cm]{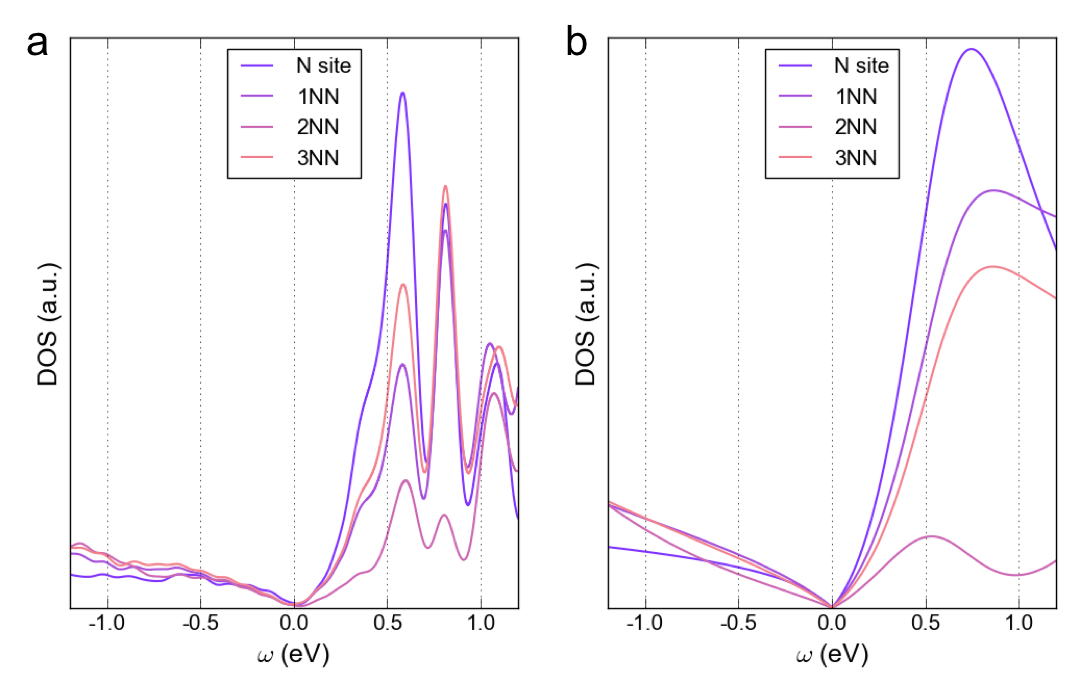}
   \caption{(a) Site-specific density of electronic states (DOS) spectra calculated using DFT and (b) respective smeared spectra.}
    \label{FigS9}
\end{figure}

\begin{figure}
    \centering
    \includegraphics[width =18cm]{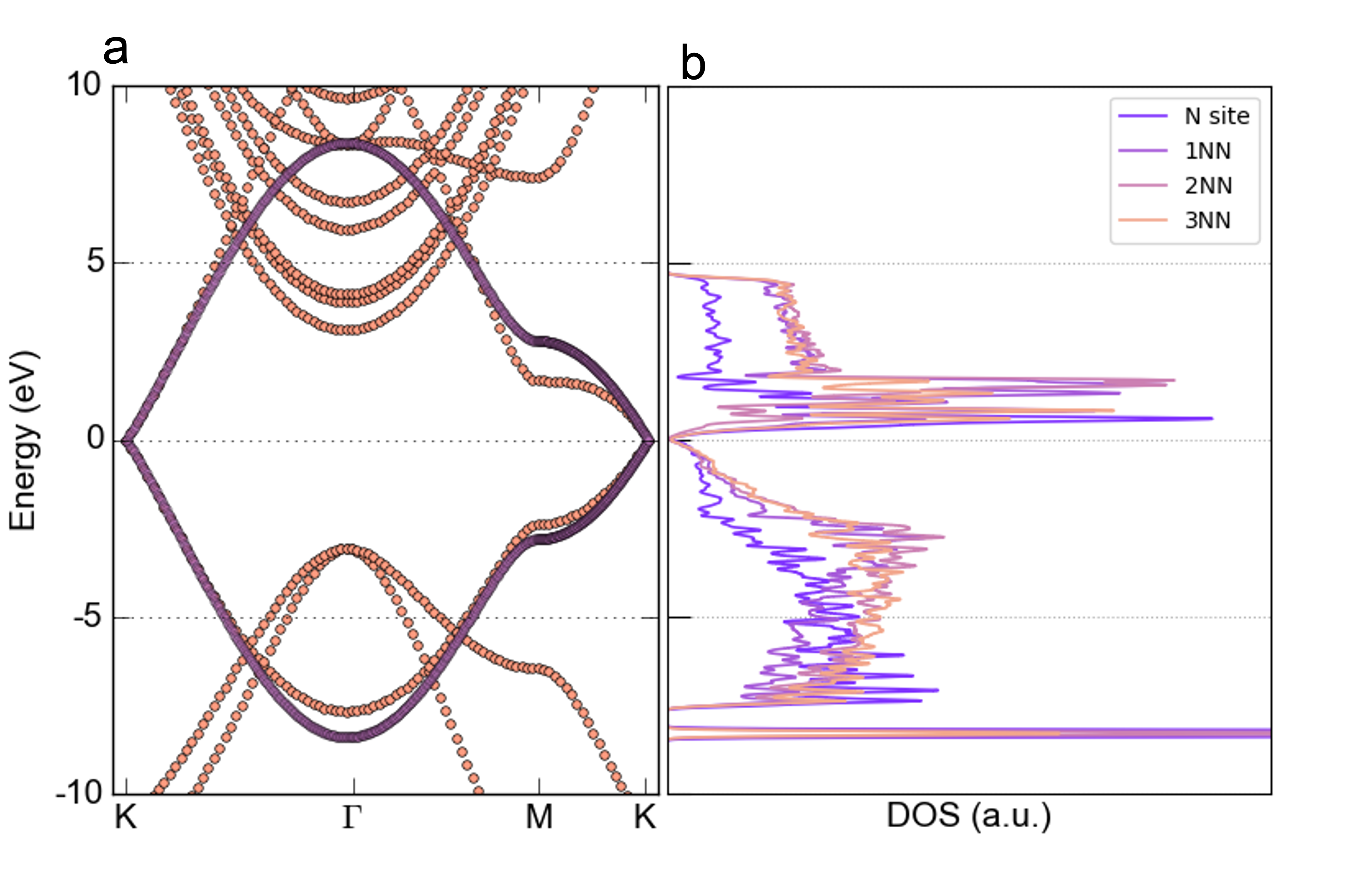}
    \caption{\textbf{Electronic structure of N-substituted graphene}. (a), Band structure of calculated using DFT of 10 $\times$ 10 unit cell. (b) DOS plot of N site, its nearest-neighbor (1NN), second nearest-neighbor (2NN) and third nearest-neighbor (3NN).}
    \label{FigS10}
\end{figure}

\begin{figure}
    \includegraphics[width = 18cm]{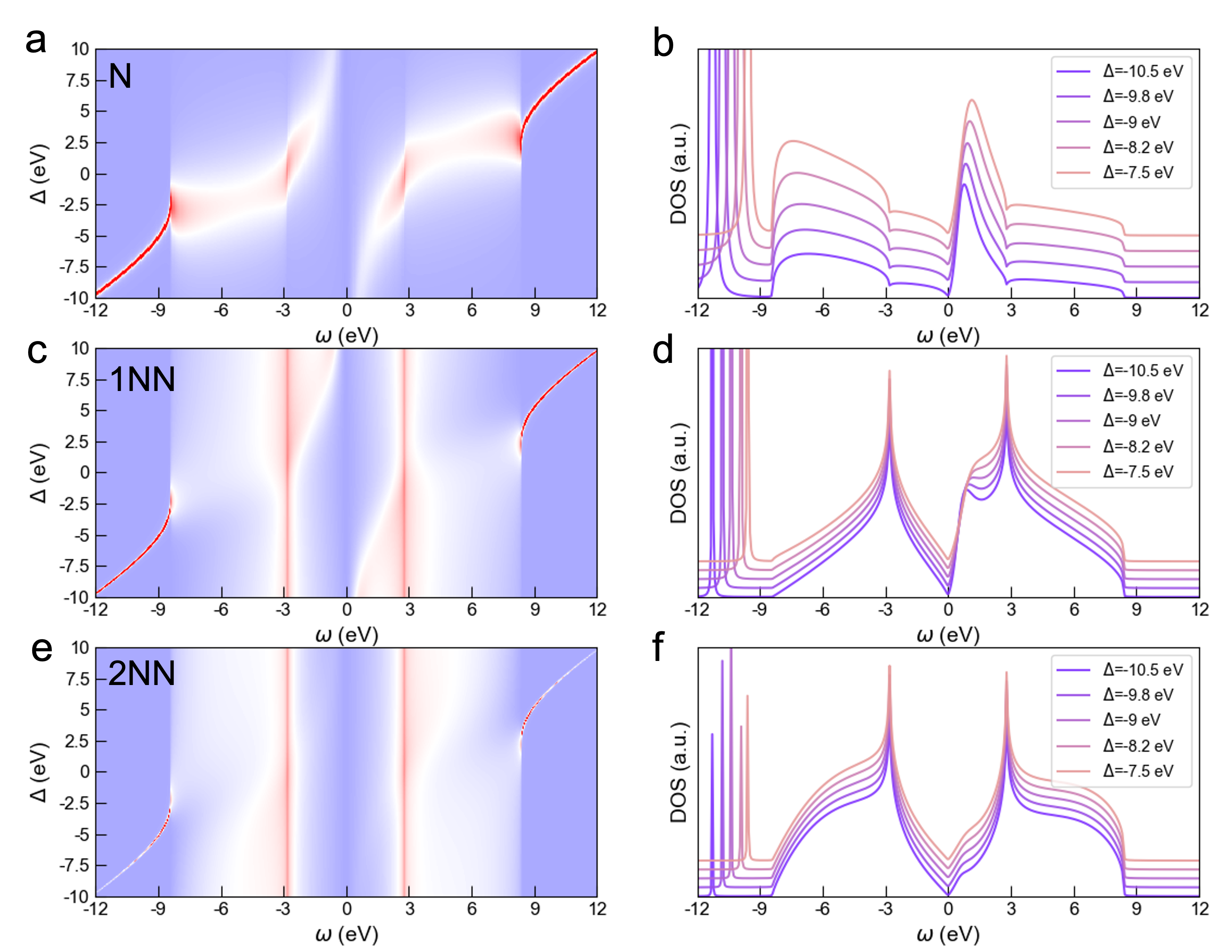}
    \caption{\textbf{Additional TB calculations}. Color-coded TB spectral plot (left) and TB spectra (right) extracted for monotonously varying values of the perturbation potential ($\Delta$) calculated for: (a,b) N site; (c,d), nearest C neighbor (1NN); (e,f) second nearest C neighbor (2NN)}
    \label{FigS11}
\end{figure}

\begin{figure}
    \centering
    \includegraphics[width = 12cm]{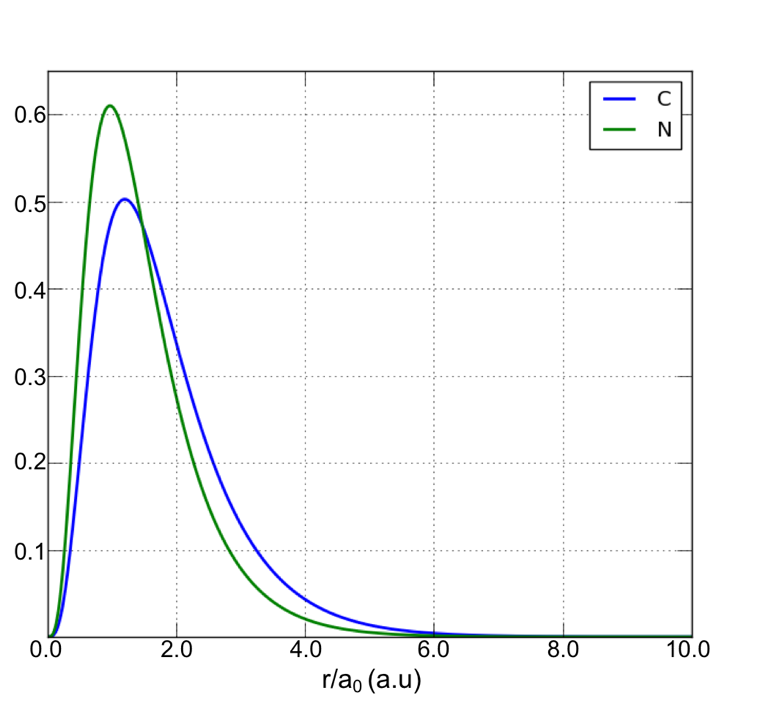}
    \caption{Radial components of the Nitrogen (green) and Carbon (blue) wave functions extracted from the DFT calculations.}
    \label{FigS8}
\end{figure}

\end{document}